\documentclass[aps,nofootinbib,twocolumn,prd,pacs]{revtex4} 
\usepackage{amsfonts}
\usepackage{amsmath}
\usepackage{graphicx}
\usepackage{epsfig}

\usepackage{pstricks}

\newcommand{\be}{\begin{equation}}
\newcommand{\ee}{\end{equation}}
\newcommand{\beq}{\begin{eqnarray}}
\newcommand{\eeq}{\end{eqnarray}}

\begin{document}
\title{Inflation in models with Conformally Coupled Scalar fields: An
  application to the Noncommutative Spectral Action} \author{Michel
  Buck\footnote{Michel.Buck@kcl.ac.uk}, Malcolm
  Fairbairn\footnote{Malcolm.Fairbairn@kcl.ac.uk} and Mairi
  Sakellariadou\footnote{Mairi.Sakellariadou@kcl.ac.uk}}
\affiliation{Department of Physics, King's College London, Strand WC2R
  2LS, London, U.K.}

\begin{abstract}
\vspace{.2cm}
\noindent
Slow-roll inflation is studied in theories where the inflaton field is
conformally coupled to the Ricci scalar. In particular, the case of
Higgs field inflation in the context of the noncommutative spectral
action is analyzed. It is shown that while the Higgs potential can
lead to the slow-roll conditions being satisfied once the running of
the self-coupling at two-loops is included, the constraints imposed
from the CMB data make the predictions of such a scenario incompatible
with the measured value of the top quark mass. We also analyze the
r\^ole of an additional conformally coupled massless scalar field,
which arises naturally in the context of noncommutative geometry, for
inflationary scenarios.
\end{abstract}

\maketitle
\section{Introduction}

Cosmological inflation is the most widely accepted mechanism to
resolve the shortcomings of the standard Hot Big Bang model. This
mechanism, leading to a phase of exponential expansion in the very
early universe, is deeply rooted in the fundamental principles of
General Relativity and Field Theory, and once combined with the
principles of Quantum Mechanics, it can account for the origin of the
observed large scale structures and the measured temperature
anisotropies of the Cosmic Microwave Background (CMB). However,
despite its success, cosmological inflation remains a paradigm in
search of a model which should be motivated by a fundamental
theory. The strength of the inflationary mechanism is based on the
assumption that its onset is generically independent of the initial
conditions. Nevertheless, even this issue is under
debate~\cite{Calzetta:1992gv,Calzetta:1992bp,Trodden:1999wc,Gibbons:2006pa,Germani:2007rt,Ashtekar:2009mm}
given the lack of a complete theory of Quantum Gravity.

The inflaton field (usually a scalar field) is assumed to dominate the
evolution of the universe at early times, but its origin and the form
of its effective potential both remain unknown; for this reason it
would be attractive if the one scalar field that {\sl is} commonly
thought to exist, namely the Higgs field, also doubled as the
long-searched for inflaton.  Unfortunately, it seems that if the Higgs
field is minimally coupled to gravity this cannot be achieved, which
has led some authors to consider large nonminimal couplings of the
Higgs field to gravity where inflation might be
achieved~\cite{Bezrukov:2007ep}.

It is commonly assumed/chosen that there is no coupling ({\sl i.e.},
minimal coupling) between the inflaton field and the background
geometry (the Ricci curvature). However, this assumption/choice seems
to lack a solid justification. A first (and merely aesthetic)
motivation comes from the observation that in the early universe
(where masses are negligible), the equations of motion for spinors and
gauge bosons have a natural conformal invariance in four space-time
dimensions, while the same is true for scalar fields only when they
couple to the Ricci scalar in a specific way.  More compelling is the
fact that even if classically the coupling between the scalar field
and the Ricci curvature could be set equal to zero, a nonminimal
coupling will be induced once quantum corrections in the classical
field theory are considered.  Moreover, a nonminimal coupling seems to
be needed in order to renormalize the scalar field theory in a curved
space-time.  The precise value of the coupling constant (denoted by
$\xi$) then depends on the choice of the theory of gravity and the
scalar field~\cite{Faraoni:1996rf}.  It has also been argued that in
all metric theories of gravity, including General Relativity, in which
the scalar field is not part of the gravitational sector ({\sl e.g.},
when the scalar field is the Higgs field), the coupling constant
should be conformal in order for the short distance propagators of the
theory to match those found in a Minkowski space-time --- a
requirement of the strong equivalence
principle~\cite{Faraoni:1996rf,Sonego:1993fw} (in our notation,
conformal coupling means $\xi=1/12$). Finally, in the context of
finite theories at one-loop level, it was
shown~\cite{Buchbinder:1989bt,Yoon:1996yr} that the nonminimal
coupling $\xi$ tends either to its conformal value or increases
exponentially in modulus, depending on the specific structure of the
theory.

In what follows, we will investigate whether scalar fields, and in
particular the Higgs field, could play the r\^ole of the inflaton in
the presence of a small positive nonminimal coupling between the
scalar field and the background geometry. The coupling constant $\xi$
is not a free parameter which could be tuned to achieve a successful
inflationary scenario avoiding severe fine-tuning of inflationary
parameters ({\sl e.g.}, the self-coupling of the inflaton field),
$\xi$ should instead be dictated by the underlying theory. For negative values of $\xi$,
exponential expansion is more easily achieved than in the minimal
case, and it can in fact lead to inflation consistent with
observational data in the strong coupling
limit~\cite{Bezrukov:2007ep,DeSimone:2008ei}.  In fact,
the slow-roll parameters for large $|\xi|$ are independent of $\xi$ and
only depend on the number of e-folds. However, exponential expansion 
is less favoured for positive values (in our conventions) such as conformal
coupling~\cite{Faraoni:2000wk}.  In light of the motivations for a
small positive $\xi$ outlined above, we will investigate whether
quantum corrections to the Higgs potential can lead to a slow-roll
inflationary era and if so, whether the constraints imposed from the
CMB temperature anisotropies are satisfied.

We will apply this analysis to the Spectral Action of NonCommutative
Geometry (NCG). This theory leads naturally to a Lagrangian with a
conformal coupling between the Higgs field and the background
geometry, in the form of a boundary condition at high energies
$E\geq\Lambda$, where $\Lambda$ is a characteristic scale of the
model. NCG provides an elegant way of accounting for the Standard
Model (SM) of Particle Physics and its
phenomenology~\cite{Chamseddine:2006ep}. Our motivation is to
investigate cosmological consequences of the NCG Spectral Action and,
in particular, to test whether slow-roll inflation driven by one of
the scalar fields arising naturally within NCG could be realized in
agreement with experimental data and astrophysical measurements.

In a previous study, we (one of us and a collaborator) have
studied~\cite{Nelson:2009wr} the conditions on the couplings so that
the Higgs field could play the r\^ole of the inflaton in the context
of the NCG.  Since however the running of couplings with the cut-off
scale had been only analyzed~\cite{Chamseddine:2006ep} neglecting the
nonminimal coupling between the Higgs field and the curvature, we were
not able to reach a definite conclusion. In this respect, the study
below is a follow-up of Ref.~\cite{Nelson:2009wr}. Moreover, it has
been argued~\cite{Marcolli:2009in} that inflation with a conformally
coupled Higgs boson could be realized in the context of NCG due to the
running of the effective gravitational constant. In what follows, we
will also analyze the validity of this statement. Finally, the NCG
Spectral Action provides, in addition to the Higgs field, another
conformally coupled (massless) scalar field, which exhibits no
coupling to the matter sector~\cite{Chamseddine:2009yf}.  One may {\sl
  a priori} wish/expect that this field could be another candidate for
the inflaton; we will examine this scenario as well.

Concluding, we analyze slow-roll inflation within models that exhibit
a conformal coupling between the Higgs field and the Ricci
curvature. Our motivation is to investigate whether any of the two
scalar fields arising naturally within the NCG Spectral Action could
be identified as the inflaton. As we will explicitly show, our
analysis leads us to the conclusion that unfortunately such a
slow-roll inflationary scenario fails to remain in agreement
with current data from high energy physics experiments and
astrophysical measurements.

This paper is organized as follows: In Section~\ref{slow-roll}, we
study the issue of the realization of slow-roll inflation within
theories with a nonminimal coupling between the scalar field and the
Ricci curvature, classically. The analysis is first performed in the
Jordan frame and then in the Einstein frame. In Section~\ref{quantum},
we consider corrections to the Higgs potential through two-loop
renormalization group analysis of the minimally coupled Standard
Model; we then enlarge this study in the case of a conformal coupling.
We focus on the gravitational and Higgs field sector of the Lagrangian
density, obtained within the noncommutative spectral action, which has
a conformal coupling, in Section~\ref{NCG}. We find that even though
we can accommodate an era of slow-roll inflation, it seems difficult
to reach an agreement with the CMB data. This conclusion holds not
only for the Higgs field but also for the other scalar field which
appears generically in the theory. We then examine in
Section~\ref{running-gravitational}, whether running of the
gravitational constant could modify our conclusions with regards to
the realization of a successful inflationary scenario driven through
one of the scalar fields in the NCG theory. We round up our
conclusions in Section~\ref{conclusions}.

Our signature convention is $(-\,+\,+\,+)$; the Riemann and Ricci
tensors are defined as
\beq
R^\sigma{}_{\mu\nu\rho}&=&\Gamma^\sigma_{\mu\rho,\nu}-\Gamma^\sigma_{\nu\rho,\mu}
+\Gamma^\tau_{\mu\rho}\Gamma^\sigma_{\tau\nu}-
\Gamma^\tau_{\nu\rho}\Gamma^\sigma_{\tau\mu}~,\nonumber\\ 
R_{\mu\nu}&=&R^\rho{}_{\mu\rho\nu}~,\nonumber
\eeq
respectively. Note that within our definition of $\xi$, conformal
coupling means $\xi=1/12$.
\section{Slow-roll inflation with Nonminimally coupled scalar fields}
\label{slow-roll} 
In this section, we will study whether slow-roll inflationary
scenarios can be realized within models with an implicit nonminimal
coupling between the inflaton field and the scalar curvature. We will
first work in the Jordan frame and then we will perform the analysis
in the Einstein frame. The Jordan frame is natural (physical) and
offers some useful insights on the effect of conformal coupling, while
the Einstein frame is mathematically more convenient, especially when
including more complicated corrections to the potential.

\subsection{Analysis in the Jordan frame}
Let us consider the action of a Higgs
boson (or any other scalar field $\phi$) nonminimally coupled to
gravity:
\be S=\int\,{\rm d}^4x\sqrt{-g}\left\{\frac{1}{2\kappa^2}f(\phi)R
-\frac{1}{2}(\nabla \phi)^2 - V(\phi)\right\}~, \ee
where 
\be
f(\phi)=1-2\kappa^2\xi\phi^2~,\nonumber
\ee
with $\kappa\equiv \sqrt{8\pi G}=m_\text{Pl}^{-1}$ and $g$ being the determinant of the
metric tensor. The scalar potential of
$\phi$ is:
\be
V(\phi)=\lambda \phi^4-\mu^2 \phi^2~.\ee
The term $-\xi \phi^2 R$ in the action encodes the explicit nonminimal
coupling of the scalar field $\phi$ to the Ricci curvature $R$.

The background geometry during inflation is of the
Fridemann-Lema\^{i}tre-Robertson-Walker (FLRW) form:
\be
{\rm d}s^2={\rm d}t^2-a^2(t){\rm d}\Sigma~,
\ee
where $t$ stands for cosmological time, $a(t)$ is the scale factor and
${\rm d}\Sigma$ describes spatial sections of constant curvature.

Einstein's equations read
\be R^{\mu\nu}-\frac12
g^{\mu\nu}R=\kappa^2f(\phi)^{-1}T^{\mu\nu}(\phi)\label{eq:EOMGrav}~, \ee
where the energy-momentum tensor, obtained by varying the action with
respect to the metric, is~\cite{Komatsu:1997hv,Tsujikawa:2004my}
\beq
T^{\mu\nu}(\phi)=\left(1-4\xi\right)\nabla^\mu\phi\nabla^\nu\phi
+4\xi\phi\left(g^{\mu\nu}\square -
\nabla^\mu\nabla^\nu\right)\phi\nonumber\\ +
g^{\mu\nu}\left[-\left(\frac12-4\xi\right)\nabla_\rho\phi\nabla^\rho\phi
  -V(\phi)\right]~.\label{eq:EMtensor}
\eeq
Here $\square\equiv g^{\mu\nu}\nabla_\mu\nabla_\nu$ is the
Laplace-Beltrami operator and Greek and Latin indices take values
0,1,2,3 and 1,2,3, respectively~\footnote{Note that it is really the
  tensor $\overline{T}^{\mu\nu}(\phi)=f(\phi)^{-1}T^{\mu\nu}(\phi)$
  which is covariantly conserved rather than
  $T^{\mu\nu}(\phi)$~\cite{Faraoni:2000wk}, but this ambiguity in the
  choice of the energy-momentum tensor will not be relevant in our
  analysis.}. The equation of motion (Klein-Gordon equation) of the
Higgs field reads
\be \square\phi-2\xi R \phi -\frac{dV}{d\phi}=0~.\label{eq:EOMHiggs}
\ee
For vanishing and quartic potentials, Eq.~(\ref{eq:EOMHiggs}) is
invariant under conformal transformations
$g_{\mu\nu}\rightarrow\Omega(x)^2g_{\mu\nu}$ and
$\phi\rightarrow\Omega(x)^{-1}\phi$ at conformal coupling $\xi=1/12$.

For a FLRW background and a spatially homogeneous $\phi$,
Eqs.~\eqref{eq:EOMGrav},\eqref{eq:EOMHiggs} combine to
\begin{align}\label{enconst}
H^2&=\frac{\kappa^2}{3f(\phi)}\left[\frac12\dot\phi^2+V(\phi)+12\xi
  H\phi\dot\phi\right],\\ 
0&=\ddot\phi+3H\dot\phi
-\frac{2\xi(1-12\xi)\kappa^2\phi\dot\phi^2}{1-2\xi(1-12\xi)\kappa^2\phi^2}~
\nonumber\\ &\quad+\frac{8\xi \kappa^2\phi
  V(\phi)+f(\phi)V'(\phi)}{1-2\xi(1-12\xi)\kappa^2\phi^2}~,
\label{fe}\end{align}
where overdots denote time derivatives and primes stand for derivatives
with respect to the argument ({\sl e.g.}, $V'(\phi)\equiv {\rm
  d}V/{\rm d}\phi$). Note that $2\xi(1-12\xi)$ is zero at both,
  minimal ({\sl i.e.}, $\xi=0$) and conformal ({\sl i.e.}, $\xi=1/12$)
  couplings.

Inflationary models are usually built upon the slow-roll
approximation, consisting of neglecting the most slowly varying terms
in the equation of motion for the inflaton field. However, in the case
of nonminimal coupling ({\sl i.e.}, $\xi\neq 0$), it is more difficult
to achieve the slow-rolling of the inflaton field. More precisely, the
nonminimal coupling term in the action, $-\xi\phi^2 R$, plays the
r\^ole of an effective mass term for the scalar field, distorting the
flatness of the scalar potential. Thus, in the case of a nonminimal
coupling, inflationary requirements such as $-\dot H < H^2$ (where $H$
denotes the Hubble parameter) do not translate in an equally
straight-forward manner to relations on the inflaton fields and their
scalar potentials.  Indeed, there is no common choice of conditions
({\sl see e.g.,}
Refs.~\cite{Komatsu:1997hv,kaiser1995primordial,Bilandzic:2007nb}),
and no analog of slow-roll parameters in terms of which quantities
such as the number of $e$-folds of expansion or perturbation
amplitudes are evaluated.

With a tentative choice of conditions~\cite{Komatsu:1997hv}
\be |\frac{\ddot\phi}{\dot\phi}|\ll H, \quad
|\frac{\dot\phi}{\phi}|\ll H \quad {\rm and} \quad
\frac12\dot\phi^2\ll V(\phi)\label{eq:slowroll}~, \ee
and a negligible mass term in the potential at high energies, the
energy constraint, Eq.~(\ref{enconst}), and field equation, 
Eq.~(\ref{fe}), reduce to
\begin{align}
H^2&\approx\frac{\lambda\kappa^2\phi^4}{3f(\phi)}
\left[1-\frac{16\xi}{1-2\xi(1-12\xi)\kappa^2\phi^2}\right]~,\\ 3H\dot\phi
&\approx-\frac{4\lambda\phi^3}{1-2\xi(1-12\xi)\kappa^2\phi^2}~,
\end{align}
respectively. These equations determine the background solution, given by
\be
a(\phi)=\left(1-2\xi\kappa^2\phi^2\right)^{\frac14}
\exp\left[-\frac{1-12\xi}{8}\kappa^2\phi^2\right]~.\label{eq:ajordan}
\ee
It is the second factor, in Eq.~(\ref{eq:ajordan}) above, which has
the potential to generate sufficient number of $e$-folds, as the first
one will only lead to logarithmic corrections. For $\xi\neq 1/12$
({\sl i.e.}, nonconformal coupling), a large enough change in
$\sqrt{|\xi|}\kappa\phi$ can lead to sufficient inflation to resolve
the horizon problem. This leaves some room to play with the coupling
and the field values, and it has indeed been
shown~\cite{DeSimone:2008ei,Bezrukov:2009db} in recent literature that
inflation can be achieved in a manner consistent with CMB data for
large negative $|\xi|\sim10^{4}$ .

At conformal coupling ($\xi =1/12$) however, the argument in the
exponential vanishes identically. For this particular value the
smallness of $\sqrt{|\xi|}$ can thus not be compensated by a larger
value of $\phi$ during inflation to generate the required
expansion.

What about quantum corrections to $\xi$? For values close to conformal
coupling, $\delta\xi=\xi-1/12$, the number of $e$-folds is
approximately
\be
N(\phi)=\frac32\delta\xi\kappa^2\left(\phi^2-\phi_{\rm e}^2\right)~,
\ee
($\phi_{\rm e}$ denotes the value of $\phi$ at the end of the
inflationary era) which requires a minimum initial Higgs field of the
order of $\phi\approx\sqrt{N/|\delta\xi|}$. Renormalization group
analysis shows that $\delta\xi$ (as a function of the energy scale) is
small in the inflationary region, namely less than
$\mathcal{O}(\xi)$~\cite{Buchbinder:1989bt,Yoon:1996yr}. The initial
Higgs amplitude required for sufficient number of $e$-folds with such
values of $\delta\xi$ generally lies above the Planck scale. Whether
this implies energies above the Planck mass relies in turn on the
value of the parameter $\lambda$. Note however that the same
renormalization group analysis of the nonminimally coupled Standard
Model suggests that there are no quantum corrections to $\xi$, if it
is exactly conformal at some energy scale
~\cite{Buchbinder:1989bt,Yoon:1996yr}. This is based on the
observation that there are no nonconformal values for the coupling
$\xi$ for which there is a renormalization group flow towards the
conformal value as one runs the Standard Model parameters up in the
energy scale. It thus indicates that if one expects an exactly
conformal coupling for the Higgs field at some specific scale, it will
be exactly conformal at all scales, hence $\delta\xi=0$.

The fact that conformal coupling destroys the accelerated expansion
has been noted previously~\cite{Faraoni:2000wk}. How can conformal
invariance be connected to the conditions for inflation? The
implications of conformal invariance on the stress-energy tensor are
well-known: if the matter sector of the theory is invariant under the
conformal transformation
\be g_{\mu\nu}\rightarrow\Omega^2g_{\mu\nu}~~~~, \qquad
\phi\rightarrow\Omega^{-1}\phi~, \ee
then the trace of $T^{\mu\nu}$ vanishes covariantly, and hence the
scalar curvature $R$ is zero. However, for a FLRW universe the scalar
curvature reads
\be
R=6(\dot H+2H^2)~,
\ee
and therefore $R=0$ implies
\be
-\frac{\dot H}{H^2}=2~,
\ee
which is, for example, satisfied during the radiation-dominated period
of the evolution of a universe in the context of General
Relativity. However, it rules out inflationary solutions which
require~\footnote{Note that conformal invariance is considered here
  solely in the matter sector.  The Einstein-Hilbert term is not
  conformally invariant.} $-\dot H/H^2<1$. Indeed, taking
$T^{\mu\nu}(\phi)$ from Eq.~\eqref{eq:EMtensor}, its trace evaluates
to
\begin{align}
\label{eqn1}
T^\mu_\mu(\phi)&=-\left[1-12\xi\right]\nabla_\rho\phi\nabla^\rho\phi
+[12\xi\phi
  V'(\phi)-4V(\phi)]\nonumber\\ &\quad + 24\xi^2R\phi^2~,
\end{align}
having used the equation of motion for the scalar,
Eq.~\eqref{eq:EOMHiggs}. However, from Eq.~\eqref{eq:EOMGrav}, the
trace of the energy-momentum thensor of $\phi$ reads
\begin{align}
\label{eqn2}
T^\mu_\mu(\phi)
 &=-\kappa^{-2}f(\phi)R\nonumber\\
&=-\kappa^{-2}(1-2\xi\kappa^2\phi^2)R~.
\end{align}
Thus, Eqs.~(\ref{eqn1}),~(\ref{eqn2}) imply
\begin{align}\label{imply}
&-\left[1-12\xi\right]\nabla_\rho\phi\nabla^\rho\phi
+[12\xi\phi
  V'(\phi)-4V(\phi)] +24\xi^2R\phi^2
\nonumber\\
&=
-\kappa^{-2}(1-2\xi\kappa^2\phi^2)R~.
\end{align}
Let us analyze Eq.~(\ref{imply}): For vanishing ($V=0$) or quartic
($V=\lambda\phi^4$) potential, conformal invariance ($\xi=1/12$)
implies that the terms in square brackets vanish and the last term on
the left-hand side cancels with the last term on the right-hand side,
leading to zero scalar curvature $R$ and thus zero trace of the
energy-momentum tensor. However, when conformal invariance is broken,
due for example to a nonzero mass term for the inflaton field ({\sl
  i.e.}, $\mu\neq0$), the induced corrections to the scalar curvature
are
\begin{align}
\delta R &= 2\mu^2\kappa^2\phi^2~.
\end{align}
In this case, the inflationary condition $-\dot H/H^2<1$ requires that
$\mu\kappa\phi>\sqrt{3}|H|$, which is not satisfied by a light scalar
inflaton.

For a $V(\phi)=\lambda\phi^4$ potential, classical analysis therefore
seems to exclude an inflationary regime. However, it is worth
investigating whether quantum corrections to the quartic self-coupling
$\lambda$ can induce potential terms that break conformal invariance,
and whether this can have a sufficiently strong effect as to enable
inflationary solutions. This can happen if these corrections are
drastic enough to generate terms in the effective potential which
alter the \emph{local} profile of the potential, {\sl i.e.},
$V(\phi)\rightarrow V_\text{eff}=V(\phi)+\alpha\delta\phi$ with
$\mathcal{O}\left((\delta\phi)'\right)\sim\mathcal{O}\left(V'\right)$. Then
the slow-roll parameters will have a different form and may allow
inflation.

For slow-roll analysis with more complex potentials, it is convenient
to perform a transformation to the Einstein frame, where the action is
formulated in terms of a rescaled metric and a new scalar field with a
minimal coupling to the curvature scalar of the new metric.  Any
meaningful conclusions should of course be independent of the choice
of conformal frame used during the calculation.

\subsection{Analysis in the Einstein Frame}

Performing a suitable Weyl transformation, the action,
Eq.~\eqref{eq:action}, can be recast in terms of a new metric 
\be
\hat g_{\mu\nu}=f(\phi)g_{\mu\nu}=\left(1-2\xi
\kappa^2\phi^2\right)g_{\mu\nu}~, \ee
and a canonical scalar field $\chi(\phi)$ that is minimally coupled and related 
to the Higgs field by 
\be \frac{{\rm d}\chi}{{\rm
    d}\phi}=\frac{\sqrt{1-2\xi(1-12\xi)\kappa^2\phi^2}}{f(\phi)}~.
\label{eq:sigmaH}
\ee
It should be noted that the transformation is singular for $\phi_{\rm
  s}=1/(\kappa\sqrt{2\xi })$. In fact, solving for the canonical field
$\chi$, one can show that it covers only the range
$|\phi|\leq\phi_{\rm s}$, implying that the analysis in the Einstein
frame is valid only for this restricted domain of the original
scalar. The value $\phi_{\rm s}$ also has special status in the Jordan
frame itself. At $\xi=1/12$ in particular, it was
shown~\cite{Starobinski:1981} that although the scalar field evolves
smoothly through $\phi_{\rm s}$ in isotropic background cosmologies,
its anisotropic shear diverges. We will safely stay below this point
in the Einstein frame analysis, still having access to Higgs field
values all the way up to the Planck scale, as long as $\xi\leq1$.

The Weyl transformation is not a diffeomorphism and the space-time
coordinates are left unchanged, $\hat x^\mu=x^\mu$. Now $a(\hat
t)=a(t)$ is not the FRWL scale factor of the universe described by the
Einstein frame variables. However, by defining a new time coordinate
\beq
{\rm d}\hat  \tau &=& a(\hat t) {\rm d}\hat t\nonumber\\
&=&a(t){\rm d}t~,
\eeq
the metric takes the FRWL form in the Einstein frame with a scale
factor
\be
\hat a(\tau)=\sqrt{f(\phi)}a(t)~,\ee
and the Hubble parameter can be defined as
\be\hat H=\frac{1}{\hat a}\frac{{\rm d}\hat a}{{\rm d}\hat \tau}~.\ee
This leaves us with an Einstein frame action
\be
S_{\rm E}=\int\,{\rm d}^4x\sqrt{-\hat g}\left\{\frac{1}{2\kappa^2}\hat R 
-\frac{1}{2}(\hat\nabla \chi)^2 - \hat V(\chi)\right\}~,
\ee
and a scalar potential
\beq \hat
V(\chi)&=&\frac{V(\phi(\chi))}{[f(\phi(\chi))]^2}\nonumber\\
&=&\frac{\lambda[\phi(\chi)]^4
-\mu^2[\phi(\chi)]^2}{[f(\phi(\chi))]^2}~.
\label{eq:Vhat}
\eeq
The expression for $\phi(\chi)$ is obtained from Eq.~(\ref{eq:sigmaH})
and can be solved analytically for any
$\xi$~\cite{Futamase:1987ua}. In this study however we shall express
any functions (\textit{e.g.}, slow-roll parameters) of the Einstein
frame in terms of $\phi$, the physical degree of freedom, so we leave
the new potential in terms of $\phi$. Of course, our interpretation of
the Einstein frame as \emph{unphysical but mathematically convenient}
presupposes that the ``observables'' computed therein have no
immediate physical meaning. We will come back to this point, and
particularly the translation from Einstein frame observables to
physical Jordan frame observables, later.\\

It is now possible to look for an inflationary regime within the
Einstein frame cosmology. We shall neglect the mass term in the
following analysis~\footnote{It is worth noting that the potential
  takes a particularly simple form at $\xi=1/12$ when the (conformal
  invariance breaking) mass term is neglected:
  $V(\chi)=36\lambda\kappa^{-4}\sinh^2{(\kappa\chi/\sqrt{6})}$.} since
we consider energy scales $E\gg\mu$. In terms of the Higgs field
$\phi$, the canonical first and second slow-roll parameters are given
by the formulae:
\begin{align}
\hat\epsilon(\phi)&= \frac{1}{2\kappa^2}\frac{1}{\hat V^2}
\left(\frac{{\rm d}\hat V}{{\rm d}\phi}\right)^2
\left(\frac{{\rm d}\chi}{{\rm d}\phi}\right)^{-2} \\
\hat\eta(\phi)&= \frac1{\kappa^2}\frac{1}{\hat V}
\left[\left(\frac{{\rm d}\chi}{{\rm d}\phi}\right)^{-2}
\frac{{\rm d}^2\hat V}{{\rm d}\phi^2}
-\frac{{\rm d}^2\chi}{{\rm d}\phi^2}
\left(\frac{{\rm d}\chi}{{\rm d}\phi}\right)^{-3}
\frac{{\rm d}\hat V}{{\rm d}\phi}\right]~.
\end{align}
The number of Einstein frame $e$-folds is
\begin{align}
\hat N&=\int_{t}^{t_\text{end}}\,\hat Hd\hat
\tau\nonumber\\ 
&=\kappa\int_{\chi_\text{end}}^{\chi}\,\frac{1}{\sqrt{2\hat\epsilon(\chi)}}
                {\rm
                  d}\chi\nonumber\\ &=\kappa\int_{\phi_\text{end}}^{\phi}\,
                \frac{1}{\sqrt{2\hat\epsilon(\phi)}}\frac{{\rm
                    d}\chi}{{\rm d} \phi} {\rm
                  d}\phi~, \label{eq:efolds}
\end{align}
and is related to the true number of $e$-folds in the Jordan frame by
\be
N=\hat N + \frac12\ln\left[\frac{f(\phi)}{f(\phi_\text{end})}\right]~. 
\label{eq:efoldrelation}
\ee
Classically, we have $\hat V(\phi)=\lambda\phi^4/[f(\phi)]^2$, which
gives
\begin{align}
\hat\epsilon(\phi)&=
\frac8{\kappa^2\phi^2[1-2\xi(1-12\xi)\kappa^2\phi^2]}\nonumber\\
&\xrightarrow{\text{CC}}\frac{8}{\kappa^2\phi^2}~,\\ 
\hat\eta(\phi)&=4\Big[\frac{3+24\xi^2\kappa^2\phi^2
-2\xi(1-12\xi)\kappa^2\left(4\xi\kappa\phi^2+1\right)\phi^2}{\kappa^2\phi^2
\left[1-2\xi(1-12\xi)\kappa^2\phi^2\right]}\Big]\nonumber\\ 
&\xrightarrow{\text{CC}}\frac43+\frac{12}{\kappa^2\phi^2}~,
\end{align}
where ${\rm CC}$ denotes the conformal coupling limit. It thus emerges
that the slow-roll parameters admit no slow-roll region at all at
conformal coupling. This can also be seen from the total number of
$e$-folds:
\be \hat N(\phi) =
\frac{(1-12\xi)\kappa^2}{8}\left(\phi^2-\phi_\text{end}^2\right)
-\frac34\ln\left[\frac{f(\phi)}{f(\phi_\text{end})}\right], \ee
which lacks the first, exponential expansion generating, term when
$\xi=1/12$. 

Comparing the number of $e$-folds in the Jordan frame, obtained in the
Einstein frame analysis, namely from Eq.~\eqref{eq:efoldrelation},
with the scalar factor $a(t)$ given from Eq.~\eqref{eq:ajordan}, one
can confirm that it indeed agrees with the previous result obtained
within the Jordan frame. This shows that the canonical slow-roll
conditions in the Einstein frame and the ones chosen in the Jordan
frame produce agreeing results, at least at the level of the observed
expansion. Of course this does not imply the equivalence of other
quantities such as the perturbation amplitudes in the two frames.  As
it has been explicitly shown in Ref.~\cite{Fakir:1991kq}, the scalar
two-point correlation functions evaluated in the Jordan frame are
different than those calculated after the field redefinitions in the
Einstein frame.  Therefore, one should keep in mind that there is a
number of ambiguities when quantum fluctuations of the scalar fields
are studied in different frames in the context of generalized Einstein
theories. Primordial spectral indices are calculated to second order
in slow-roll parameters in Ref.~\cite{kaiser1995primordial} for
different inflationary models, in the context of theories with a
nonminimal coupling between the inflaton field and the Ricci curvature
scalar. It has been shown that there are inflationary models ({\sl
  e.g.}, new inflation) for which there are discrepancies between the
values of the spectral index $n_{\rm s}$ calculated in the Einstein
and the Jordan frame, while for others ({\sl e.g.}, chaotic inflation)
there are not. Finally, the reader should keep in mind that while
realization of slow-roll inflation in the (physical) Jordan frame, in
which the inflaton is nonminimally coupled to the Ricci curvature,
implies slow-roll inflation in the (unphysical) Einstein frame, the
{\sl vive versa} does not hold~\cite{Faraoni:2000wk}.

\section{Flat potential through quantum corrections}
\label{quantum}
The Higgs potential takes the classical form
\be
V(\phi)=\lambda \phi^4-\mu^2 \phi^2~,
\ee
however both $\mu$ and $\lambda$ are subject to radiative corrections
as a function of energy.  For very large values of the field $\phi$
one therefore needs to calculate the renormalized value of these
parameters at the energy scale $\mu\sim\phi$.  The running of the top
Yukawa coupling and the gauge couplings cannot be neglected and must
be evolved simultaneously. We follow the analysis of
Ref.~\cite{Espinosa:2007qp}, which relies upon the $\beta$-functions and
improved effective potential presented in
Refs.~\cite{Kastening:1991gv,Ford:1992mv,Bando:1992wz}.  This involves
taking the measured values of the gauge couplings at low energy and
evolving them upwards in energy, taking into account the thresholds
where quark species come into the running.  It is neccesary to
simultaenously evolve all three gauge couplings and the top quark
Yukawa coupling in order to accurately predict the full effect upon
the Higgs self-coupling.  Care must be taken to use the correct
relationship between the pole masses and the parameters used in the
running \cite{Espinosa:2007qp}.

At high energies the mass term is sub-dominant and one can write the
effective potential as
\be
V(\phi)=\lambda(\phi) \phi^4~.
\ee
Then for a given mass $m_{\rm t}$ of the top quark, a smaller value of
the Higgs mass will result in the quartic coupling being driven down
at large values of $\phi$, such that it may develop a metastable or
true vacuum at expectation values of $\phi$, far in excess of that
observed from Standard Model physics $\left<\phi\right>=246 {\rm
  GeV}$.  For typical values of $m_{\rm t}$, if this false vacuum
appears at all, it will show up relatively close to the Planck scale.
When calculating the running of $\lambda$ it is in fact necessary to
go to two-loop accuracy since at one-loop this second minimum develops
at scales typically far in excess of the Planck scale, where we would
really expect higher order nonrenormalizable contributions to the
potential to become important.

\begin{figure}[h]
\centering
\includegraphics[width=0.48\textwidth]{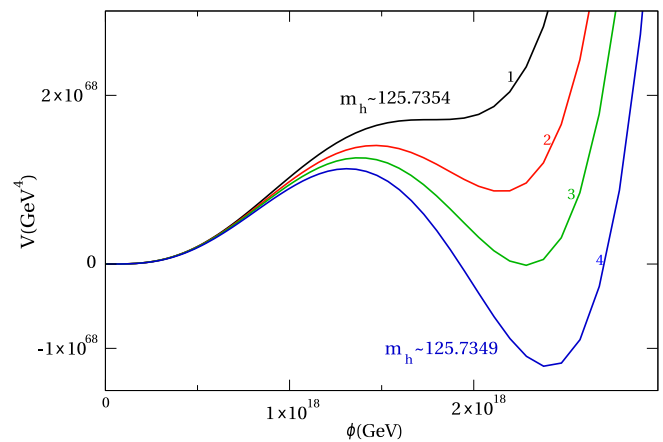}
\caption{Sub-Planckian flattening of the Higgs potential due to
  two-loop corrections in the Standard Model ($\xi=0$). We analyze
  slow-roll for profiles just above the top (black) curve, which
  feature no metastable vacua.}
\label{fig:potential}
\end{figure}

\begin{figure}[h]
\centering
\includegraphics[width=0.4\textwidth]{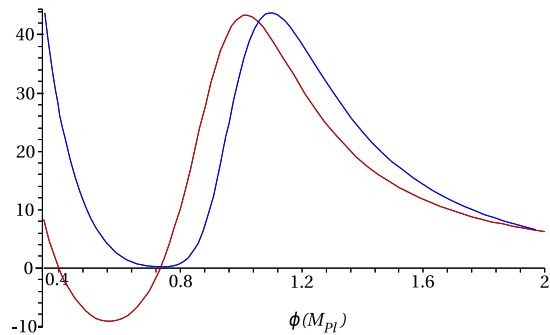}
\caption{Typical profiles of $\epsilon$ (blue) and $\eta$ (red) with a
  small sub-Planckian region of slow-roll, plotted here for $m_{\rm
    t}=172 {\rm GeV}$ and $\delta=0$. There is a narrow region in
  which both are very small.}
\label{fig:etaepsilon}
\end{figure}

For each value of $m_{\rm t}$ there is therefore a value of the Higgs
mass, $m_{\rm h }$, where the effective potential is on the verge of
developing a metastable minimum at large values of $\phi$ and the
Higgs potential is locally flattened. This is
illustrated in Fig.~\ref{fig:potential}. Since the region where the
potential becomes flat is narrow, slow-roll must be \emph{very} slow
({\sl i.e.}, the slow-roll parameters very small), in order to provide a
sufficiently long period of quasi-exponential expansion. The slow-roll
parameters for the top (black curve) potential profile of
Fig.~\ref{fig:potential} are shown in Fig.~\ref{fig:etaepsilon}, and
one can see that the region where $\epsilon$ is extremely small takes
the form of a narrow dip. It is there that the integral $N\sim\int
\epsilon^{-\frac12} {\rm d}\phi\,$ can generate the required number of
$e$-folds. \\

It was noted in Ref.~\cite{Isidori:2007vm} that in the minimally
coupled model, slow-roll through this flat region will not match the
observed amplitude of density perturbations $\Delta_\mathcal{R}^2$ in
the cosmic microwave background. Inflation predicts the latter to be
related to the potential and first-slow roll parameter at horizon
crossing (labelled by stars). Its value as measured by WMAP7
~\cite{Larson:2010gs} imposes the constraint
\be \left(\frac{V_*}{\epsilon_*}\right)^{\frac14}
=2\sqrt{3\pi}m_\text{Pl}\Delta_\mathcal{R}^\frac12=(2.75\pm0.30)
\times10^{-2}m_\text{Pl}\,~,\label{eq:cobe} \ee
where $\epsilon_*\leq1$. The mismatch arises because $\epsilon$ needs
to be extremely small in order to allow for sufficient $e$-folds and
the potential energy is then too large to fit the condition. However,
even in the minimally coupled model, there remains the possibility
that horizon crossing occurs close to the beginning of inflation,
where $\epsilon$ is not yet so small, provided the flat region occurs
at low enough energy. Since $\epsilon_*\leq1$, the maximum potential
energy at horizon crossing is $5.7\times10^{-7}m_\text{Pl}^4$. We
shall see in the renormalization group analysis that there exist
values of the top quark mass for which the flattening does happen at
energies below this value. Furthermore, the presence of nonminimal
coupling has additional effects since it changes the potential felt by
the Higgs field.

When the nonminimal coupling $\xi$ of the Higgs boson to gravity is
included in the Standard Model, it has a $\beta$-function induced by
the coupling between the Higgs field and the matter sector whose behaviour
has been analyzed to one-loop~\cite{Buchbinder:1989bt,Yoon:1996yr}. As
previously stated, we take $\beta_\xi=0$, since the presence of a
boundary value $\xi=1/12$ at some energy scale suggests that
$\xi=1/12$ at all scales. The $\beta$-function of the quartic Higgs
self-coupling changes as well due to the $-\xi R\phi^2$ term, and this
can have significant effects on the remaining Standard Model
parameters when $\xi$ is large~\cite{Lerner:2009xg}. We have worked
out how large $\xi$ needs to be to impact the normal Standard Model
running by considering the two cases $\xi=1$ and $\xi=-1$ at low
energies and running these up with the other parameters.  The effect
that either of these choices has on the potential is very small and
looks like a minute change in the Higgs mass, much less than than any
possible experimental error.  Because we are well within this range we
can neglect these corrections.

We therefore calculate the renormalization of the Higgs self-coupling
in the minimally coupled Standard Model and construct an effective
potential which fits the renormalization group improved potential
around the flat region. The modifications in that fit are very small
when the conformal coupling is included. We first consider the
implications for the minimally coupled model (where the Jordan and
Einstein frames coincide), which had been mentioned in
Ref.~\cite{Isidori:2007vm}, and then extend the analysis to a
conformally coupled model, which is of particular relevance to the
Noncommutative Spectral Action approach to the Standard Model of
Particle Physics.

There is a very good analytic fit to the Higgs potential in the region
around this plateau/minimum, which takes the form
\beq
V_\text{E}&=&\lambda_\text{E}(\phi)\phi^4\nonumber\\
&=&[a\ln^2(b\kappa\phi)+c]\phi^4~.
\eeq
The parameters are found to relate to the low energy values of
$m_{\rm t}$ in the following way:
\begin{align}
a(m_\text{t})&=4.04704\times10^{-3}-4.41909\times10^{-5}
\left(\frac{m_\text{t}}{\text{GeV}}\right)\nonumber\\ &\quad
+1.24732\times10^{-7}\left(\frac{m_\text{t}}{\text{GeV}}\right)^2~,
\nonumber\\ 
b(m_\text{t})&=\exp{\left[-0.979261
\left(\frac{m_\text{t}}{\text{GeV}}-172.051\right)\right]}~.
\end{align}
The third parameter, $c=c(m_\text{t},m_\phi)$, encodes the appearance
of an extremum ({\sl see}, Fig.~\ref{fig:potential}) and depends on
the values for $m_\text{t}$ and $m_\phi$. Indeed, $V_\text{E}(\phi)$
exhibits a sub-Planckian flat region (or local minimum) for suitably
tuned parameters. An extremum occurs if and only if $c/a\leq 1/16$,
the saturation of the bound corresponding to a perfectly flat region,
{\sl i.e.}, $V_\text{E}'(\phi_0)=V_\text{E}''(\phi_0)=0$, where
$\phi_0=e^{-\frac14}/b(m_{\rm t})$ and $e$ is Euler's constant. The energy
at these points is given by
\be
V_\text{E}(\phi_0)=\frac{a(m_\text{t})}{8eb(m_\text{t})^4}\kappa^4,
\ee
which for $169\leq m_\text{t}\leq175$ lies within
$10^{-10}\kappa^{-4}\leq V_0 \leq \kappa^{-4}$ (note that $V_{\rm
  E}(\phi_0)$ increases with $m_{\rm t}$). This shows that there are regions
where the flattening occurs at scales potentially consistent with
perturbation amplitudes, given in Eq.~\eqref{eq:cobe}. 

It is convenient to write $c=[(1+\delta)/16]a$, where $\delta=0$
saturates the bound below which a local minimum is formed.  We
restrict ourselves to $\delta>0$, so that the potential contains no
metastable vacua.  The slow-roll conditions are met only for a narrow
region, but for the points in parameter space which are close to
$\delta=0$, both slow-roll parameters vanish simultaneously and we get
slow-roll inflation with extremely small $\epsilon$.
From Eq.~\eqref{eq:cobe} it follows that for $m_\text{t}>171.42
\text{GeV}$ the two conditions cannot be simultaneously met since the
flat region occurs at too high energies. Slow-roll is restricted to
the domain where $\max\left(\epsilon,|\eta|\right)\leq1$, and for
inflation one should find a point in parameter space which: (i) leads
to sufficient $e$-folds within a region
$\left[\phi_\text{end},\phi_*\right]$, (ii) has an $\epsilon_*$ which
lies within the bounds imposed by COBE normalization, and (iii)
satisfies the observational constraints on $n_{\rm s}$ and $r$. The
measured value of perturbation amplitudes serves as a convenient first
test of the model. For the scenario to be viable, $\epsilon$ at
horizon crossing cannot be too small. Since the requirement on a
sufficient number of $e$-folds relies on a potential that has very
small $\epsilon$ in a small region, the problem is that the valley in
$\epsilon$ is far narrower than that in $\eta$. As a result, within
the region $|\eta|\leq1$, $\epsilon$ tends to be very small.  The best
fit to the observed perturbation amplitude will occur for scenarios in
which horizon crossing occurs close to the onset of inflation,
{\sl i.e.}, $\eta(\phi_\star)\sim1$, so that $\epsilon_\star$ takes its
largest possible value.

The corrections due to conformal coupling to the potential in the
Einstein frame are entirely embodied in the function
$f(\phi)\sim1+\mathcal{O}(\kappa^2\phi^2)$, since the canonical field
$\chi$ feels the potential $V_\text{E}/f^2$.  The value of the Higgs
field where the plateau occurs in the potential rises with increasing
top quark mass, so the greatest effect will be at the highest top
quark mass. However, the lower bound on $\epsilon_*$ then gets more
stringent since $V_*$ is larger. Due to the change in the potential,
flatness does not occur at $\delta=0$ anymore but for fixed values of
$\delta$ depending on the value of the top quark mass. Sub-Planckian
inflation is again reliant on a relationship between the Higgs field
and the top quark masses. The values of $\delta$ for which the
potential has the right flatness are not anymore centered around
$\delta=0$ due to the altered form of the potential. This has an
effect on the Higgs masses where flattening occurs: for any
$-1\leq\delta\leq0$, a given top quark mass fixes the Higgs mass to a
value in the range $(120-130)\,\text{GeV}$ with an accuracy of $\Delta
m_\phi/m_\phi\sim10^{-6}$. This means that for inflation to occur via
this mechanism, the top quark mass fixes the Higgs mass extremely
accurately. As an example, for $m_\text{t}=171.70\, \text{GeV}$ and
$\delta=-0.2867$ (corresponding to $m_\phi=125.735368\, \text{GeV}$),
we obtain $\hat N=62$ of $e$-folds between $\kappa\phi=0.9570$ and
$\kappa\phi_\text{end}=0.9417$.\\

Scanning through parameter space it emerges that sufficient $e$-folds
are indeed generated provided a suitably tuned relationship between
$m_{\rm t}$ and $m_\phi$ holds. Numerical integration needs to be
performed carefully since the slow-roll approximation implies a
strongly peaked integrand in the number of $e$-folds. Using a
Runge-Kutta integrator in FORTRAN we identify the curve in parameter
space along which sufficient expansion occurs during almost perfect
de\,Sitter inflation, both for minimal and conformal couplings. 

The next step is a comparison with astrophysical measurements. To
probe the parameter space more finely we use a Monte-Carlo chain. It
turns out that in both the minimal and conformal cases, the
perturbation amplitudes are too large --- the best fit to the ratio
$(V_\star/\epsilon_\star)^{\frac14}$ is still too large by two orders
of magnitude. Small positive nonminimal couplings such as $\xi=1/12$
improve the fit only minimally.  It should be noted also that when
perturbation amplitudes are too large, scenarios where perturbations
are generated by a curvaton are in turn ruled out as well, because the
quantum fluctuations of the inflaton are already too large.

\begin{figure}[h]
\centering
\includegraphics[width=0.48\textwidth]{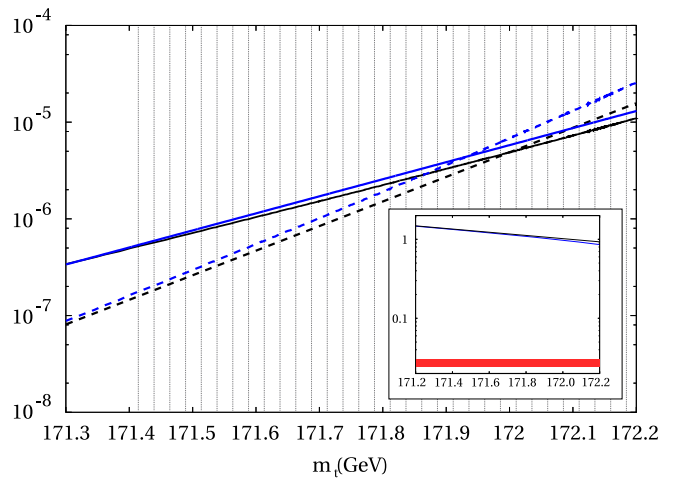}
\caption{The value of the potential (solid) in units of $\kappa^{-4}$
  and the maximum value of the first slow-roll parameter (dashed) at
  horizon crossing for minimal $\xi=0$ (black) and conformal
  $\xi=1/12$ (blue). The striped area represents the region of the top
  mass excluded by Eq.~\eqref{eq:cobe} from the height of the plateau
  in the potential. The inset shows the ratio
  $(V_\star/\epsilon_\star)^{\frac14}$ in both cases and WMAP7
  observations (red region). The calculated value of perturbation
  amplitudes is off by several orders of magnitude and the improvement
  at conformal coupling minimal.}
\label{fig:epshigh}
\end{figure}

In Fig.~\ref{fig:epshigh} we show the best fit, {\sl i.e.}, the
scenario with the largest possible values of $\epsilon_*$, the first
slow-roll parameter at horizon crossing, for a given top quark mass
along with the potential energy $V_*$, at horizon crossing.  The
resulting ratio of perturbation amplitudes is too large for any value
of $m_{\rm t}$.

On a side note, let us mention that the renormalization of Standard
Model parameters is generally performed in Minkowski space-time, while
inflationary perturbations are calculated on a general de\,Sitter
background. The conditions for the geralization of a slow-roll
inflationary era should of course be studied in a de\,Sitter space and
the Coleman-Weinberg result should then be recovered as a limit to the
flat Minkowski space-time. This analysis has been performed in a
recent study~\cite{Bilandzic:2007nb}, where the one-loop
improved potential for the nonminimally coupled scalar $\lambda\phi^4$
theory in de Sitter space was calculated. Their analysis poses a
stringent constraint on the coupling parameter $\xi$. The assumption
$|\dot H|< H^2$ along with the requirement that $f(\phi)$ in the
equations of motion remain nonsingular, $f(\phi)<\infty$, implies
\be
\frac1{16\tilde{N}}\ll|\xi|\ll\frac1{48}~,
\label{eq:prokopec}
\ee where $\tilde{N}\approx N+1-\xi$. This rules out most values of
$\xi$ used in literature.  However, $|\dot H|< H^2$ is in fact a
stronger condition than the condition $-\dot H < H^2$ for inflationary
expansion. The latter implies the former only when $\dot H$ is
negative. The stronger condition $|\dot H|< H^2$ could be circumvented
in an inflationary universe where $\dot H$ is large and positive.  In
the minimally coupled case this is clearly not possible, since $\dot H
= -\dot \phi^2/2$. For nonzero $\xi$ we have however
\be \dot H=\frac{\phi\dot\phi}{f(\phi)}\left[-\frac12
  \left(1-4\xi\right)\frac{\dot\phi}{\phi} -2\xi H+2\xi
  \frac{\ddot\phi}{\dot\phi}\right]~,
\label{conditionH}
\ee
wich for the slow-roll conditions, Eq.~(\ref{eq:slowroll}), reduces to
\be \dot H =-\frac{2\xi\phi\dot\phi}{f(\phi)}H~.  \ee
Since the field rolls down the potential,
$\rm{sign}(\phi)=-\rm{sign}(\dot\phi)$ and $\dot H$ is indeed positive
when the nonminimal coupling is positive ({\sl e.g.}, conformal) in
our notation. This means that the above constraint does not apply in
the conformally coupled case. However, it should be mentioned that for
negative choices of $\xi$, popular due to their promise in achieving
Higgs driven inflation, $\dot H < 0$. The constraint in
Eq.~\eqref{eq:prokopec} is then valid and seems to be in contradiction
with large $|\xi|$.

\section{Noncommutative Spectral Action and Inflation}
\label{NCG}

Using the language of noncommutative geometry and spectral triples,
Connes and collaborators have reformulated the Standard Model in terms
of purely geometric data~\cite{Chamseddine:2006ep}.  Based on spectral
triples, A.~Connes~\cite{Connes:1994yd} has developed a new calculus
that deals not with the underlying spaces, but with the algebra of
functions defined upon them instead. This reformulation allows a
natural generalization of the differential calculus on Riemannian
manifolds to a wider class of geometric structures, {\sl i.e.},
noncommutative spaces. It is the geometry of these spaces that encodes
not only space-time and gravity, but also the matter content of the
Standard Model.

In NCG, the fundamental particles and interactions derive from the
spectral data of an action functional defined on noncommutative
spaces, the Spectral Action. The Standard Model emerges as the
asymptotic expansion of this action at an energy $\Lambda$ below the
Planck scale, at which the fundamental noncommutative space is
approximated by an almost-commutative space. This space is assumed to
be the simplest noncommutative extension of the smooth
four-dimensional space-time manifold, and is obtained by taking its
tensor product with a finite noncommutative space.  Having recovered
low energy physics in the framework of NCG, the next step will be to
find the true geometry at Planckian energies, for which this product
is a low energy limit. We consider here the effective action
functional at the scale $\Lambda$.

In this section, we will first highlight the main principles of the
noncommutative geometry approach and we will then investigate possible
inflationary mechanisms driven by one of the available scalar fields.

\subsection{Elements of NCG spectral action}

Within General Relativity, the group of symmetries of gravity is given
by diffeomorphism of the underlying differentiable manifold of
space-time; a key ingredient that one would like to extend to the
theory of elementary particles. To achieve such a {\sl geometrization}
of the Standard Model coupled to gravity, one should turn the SM
coupled to gravity into pure gravity on a preferred space, whose group
of diffeomorphisms is given by the semidirect product of the group of
maps from the background manifold to the gauge group of the SM, with
the group of diffeomorphisms of the background manifold. Such
preferred space cannot be obtained however within ordinary spaces,
while noncommutative spaces can easily lead to the desired
answer. This is the main reason for extending the framework of
geometry to spaces whose algebra of coordinates is noncommutative.

To extend the Riemaniann paradigm of geometry to the notion of metric
on a noncommutative space, the latter should contain the Riemaniann
manifold with the metric tensor (as a special case), allow for
departures from commutativity of coordinates as well as for quantum
corrections of geometry, contain spaces of complex dimension, and
offer the means of expressing the Standard Model coupled to Einstein
gravity as pure gravity on a suitable geometry.  A metric NCG is given
by a spectral triple $({\cal A}, {\cal H}, D)$, in the sense that we
will discuss below.  Thus, within NCG, geometric spaces emerge
naturally from purely spectral data. The fermions of the Standard
Model provide the Hilbert space ${\cal H}$ of a spectral triple for a
suitable algebra ${\cal A}$, and the bosons arise naturally as inner
fluctuations of the corresponding Dirac operator $D$. To study the
implications of this noncommutative approach coupled to gravity for
the cosmological models of the early universe, we will only consider
the bosonic part of the action; the fermionic part is however crucial
for the particle physics phenomenology of the model.

More precisely, let us consider a geometric space defined by the
product of a continuum compact Riemaniann manifold, ${\cal M}$, and a
tiny discrete finite noncommutative space, ${\cal F}$, composed of
only two points.  The product geometry ${\cal M}\times{\cal F}$ has
the same dimension as the ordinary space-time manifold, namely
4. Hence, the noncommutative space ${\cal F}$ has zero metric
dimension.  The space ${\cal F}$ represents the geometric origin of
the Standard Model and it is specified in terms of a real spectral
triple $({\cal A},{\cal H},D)$, where ${\cal A}$ is a noncommutative
$^\star$-algebra, ${\cal H}$ is a Hilbert space on which ${\cal A}$ is
realized as an algebra of bounded operators, and $D$ is a suitably
defined Dirac operator on ${\cal H}$.  The Dirac operator can be seen
as the inverse of the Euclidean propagator of fermions. Since the
action functional only depends on the spectrum of the line element, it
is a purely gravitational action. In other words, the physical
Lagrangian is entirely determined by the geometric input, which
implies that the physical implications are closely dependent on the
underlying chosen geometry, {\sl see}, Ref.~\cite{Chamseddine:2006ep}.

By assuming that the algebra constructed in ${\cal M}\times {\cal F}$
is {\it symplectic-unitary}, the algebra ${\cal A}$ is restricted to
be of the form
\begin{equation}
\mathcal{A}=M_{a}(\mathbb{H})\oplus M_{k}(\mathbb{C})~,
\end{equation}
where $k=2a$ and $\mathbb{H}$ is the algebra of quaternions.  The
choice $k=4$ is the first value that produces the correct number
($k^2=16$) of fermions in each of the three
generations~\cite{Chamseddine:2007ia}.  The Dirac operator $D$
connects ${\cal M}$ and ${\cal F}$ via the spectral action functional
on the spectral triple.  It is defined as ${\rm Tr}\left(
f\left(D/\Lambda\right)\right)$, where $f>0$ is a test function and
$\Lambda$ is the cut-off energy scale.  The asymptotic expression for
the spectral action, for large energy $\Lambda$, is of the form
\begin{equation}
\label{eq:sp-act}
{\rm Tr}\left(f\left({D\over\Lambda}\right)\right)\sim 
\sum_{k\in {\rm DimSp}} f_{k} 
\Lambda^k{\int\!\!\!\!\!\!-} |D|^{-k} + f(0) \zeta_D(0)+ {\cal O}(1)~,
\end{equation}
where $f_k= \int_0^\infty f(v) v^{k-1} {\rm d}v$ are the momenta of
the function $f$, the noncommutative integration is defined in terms
of residues of zeta functions, and the sum is over points in the {\sl
  dimension spectrum} of the spectral triple.  The test function
enters through its momenta $f_0, f_2, f_4$; these three additional
real parameters are physically related to the coupling constants at
unification, the gravitational constant and the cosmological constant.
In the four-dimensional case, the term in $\Lambda^4$ in the spectral
action, Eq.~(\ref{eq:sp-act}), gives a cosmological term, the term in
$\Lambda^2$ gives the Einstein-Hilbert action functional with the
physical sign for the Euclidean functional integral (provided
$f_2>0$), and the $\Lambda$-independent term yields the Yang-Mills
action for the gauge fields corresponding to the internal degrees of
freedom of the metric. The scale-independent terms in the spectral
action have conformal invariance. Note that the arbitrary mass scale
$\Lambda$ can be made dynamical by introducing a scaling dilaton
field.

Writing the asymptotic expansion of the spectral action, a number of
geometric parameters appear; they describe the possible choices of
Dirac operators on the finite noncommutative space. These parameters
correspond to the Yukawa parameters of the particle physics model and
the Majorana terms for the right-handed neutrinos.  The Yukawa
parameters run with the renormalization group equations of the
particle physics model. Since running towards lower energies, implies
that nonperturbative effects in the spectral action cannot be any
longer safely neglected, any results based on the asymptotic expansion
and on renormalization group analysis can only hold for early universe
cosmology. For later times, one should instead consider the full
spectral action.

Applying the asymptotic expansion of Eq.~(\ref{eq:sp-act}) to the
spectral action of the product geometry ${\cal M}\times{\cal F}$ gives
a bosonic functional $S$ which includes cosmological terms, Riemannian
curvature terms, Higgs minimal coupling, Higgs mass terms, Higgs
quartic potential and Yang-Mills terms.  Moreover, one can introduce a
relation between the parameters of the model, namely a relation
between the coupling constants at unification. More precisely, we
impose the relation
\be {g_3^2f_0\over 2\pi^2}={1\over 4} ~~\mbox{and}~~
g_3^2=g_2^2={5\over 3}g_1^2~, \ee
between the coefficient $f_0$ and the coupling constants $g_1, g_2,
g_3$, which is dictated by the normalization of the kinetic
terms. This condition means that the so-obtained spectral action has
to be considered as the {\sl bare action} at unification scale $\Lambda$,
where one supposes the merging of the coupling constants to take
place.

The gravitational terms in the spectral action, in Euclidean
signature, are of the form
\beq\label{eq:spectral-action}{\cal S}_{\rm grav}^{\rm E} = \int \left(
\frac{1}{2\kappa^2} R + \alpha_0
C_{\mu\nu\rho\sigma}C^{\mu\nu\rho\sigma} + \tau_0 R^\star
R^\star\right.  \nonumber\\ -\left.  \xi_0 R|{\bf H}|^2 \right)
\sqrt{g} {\rm d}^4 x~. \eeq
Note that ${\bf H}$ is a rescaling ${\bf H}=(\sqrt{af_0}/\pi)\phi$ of
the Higgs field $\phi$ to normalize the kinetic energy; the momentum
$f_0$ is physically related to the coupling constants at unification
and the coefficient $a$ is related to the fermion and lepton masses
and lepton mixing.  In the above action,
Eq.~(\ref{eq:spectral-action}), the first two terms only depend upon
the Riemann curvature tensor; the first is the Einstein-Hilbert term
with the second one being the Weyl curvature term. The third term
\be R^\star
R^\star=\frac{1}{4}\epsilon^{\mu\nu\rho\sigma}\epsilon_{\alpha\beta\gamma\delta}
R^{\alpha\beta}_{\mu\nu}R^{\gamma\delta}_{\rho\sigma}~,\nonumber\ee
is the topological term that integrates to the Euler characteristic
and hence is nondynamical. Notice the absence of quadratic terms in
the curvature; there is only the term quadratic in the Weyl curvature
and topological term $R^\star R^\star$. In a cosmological setting
namely for Friedmann-Lema\^{i}tre-Robertson-Walker geometries, the Weyl
term vanishes. The spectral action contains one more term that couples
gravity with the SM, namely the last term in
Eq.~(\ref{eq:spectral-action}), which should always be present when
one considers gravity coupled to scalar fields.

\subsection{Higgs field inflation}
The asymptotic expansion of the Spectral Action, proposed in
Ref.~\cite{Chamseddine:2006ep}, gives rise to the following
Gravity-Higgs sector $\mathcal{L}_\text{GH}\subset
\mathcal{L}_\text{NCG}$:
\be
S_\text{GH}=\int\,d^4x\sqrt{-g}\left\{\frac{1-2\kappa^2\xi{\bf H}^2}{2\kappa^2}R
-\frac{1}{2}(\nabla {\bf H})^2- V({\bf H})\right\}\label{eq:action}~, \ee
where $V({\bf H})=\lambda{\bf H}^4-\mu^2{\bf H}^2$. In the derivation
of the Standard Model from the Spectral Action principle, the metric
carries Euclidean signature. The discussion of phenomenological
aspects of the theory relies on a Wick rotation to imaginary time,
into the standard (Lorentzian) signature. While sensible from the
phenomenological point of view, there exists as yet no justification
on the level of the underlying theory.

To discuss the phenomenology of the aspects of the cut-off scale
$\Lambda$, the Spectral Action principle leads to a number of boundary
conditions on the parameters of the Lagrangian. These conditions
encode the geometric origin of the Standard Model
parameters. Normalization of the kinetic terms in the action implies
the following relations:
\beq
\kappa^2&=&\frac{12\pi^2}{96f_2\Lambda^2-f_0\mathfrak{c}}
~,\nonumber\\
 \xi &=&\frac{1}{12}~,\nonumber\\
 \lambda&=&\frac{\pi^2\mathfrak{b}}{2f_0\mathfrak{a}^2}~,\nonumber\\
\mu &=&2\Lambda^2\frac{f_2}{f_0}~.
\eeq
We emphasize that the action, Eq.~(\ref{eq:action}), has to be
taken as the {\sl bare action} at some cutoff scale $\Lambda$. The
renormalized action will have the same form but with the bare
quantities $\kappa, \mu, \lambda$ and the three gauge couplings
$g_1,g_2,g_3$ replaced with physical quantities.

The factor $f_0$ is fixed by the canonical normalization of the
Yang-Mills terms (not included here) in terms of the common value of
the gauge coupling constants $g$ at unification,
$f_0=\pi^2/(2g^2)$. The value of $g$ at the unification scale is
determined by standard renormalization group flow, {\sl i.e.}, it is
given a value which reproduces the correct observed coupling at low
energies.  Note that it is not unique since the gauge couplings fail
to meet exactly in the nonsupersymmetric Standard Model (or its
extension by right-handed neutrinos). The coefficients
$\mathfrak{a},\mathfrak{b},\mathfrak{c}$ are the Yukawa and Majorana
parameters subject to renormalization group flow, see {\sl e.g.}
Ref.~\cite{Chamseddine:2006ep}. The parameter $f_2$ is {\sl a priori}
unconstrained in $\mathbb{R}^*_+$.

Assuming the {\sl big desert} hypothesis, we can connect the physics
at low energies with those at $E=\Lambda$ through the standard
renormalization procedure.  This was carried out at one loop in
Ref.~\cite{Chamseddine:2006ep}, and more recently in
Ref.~\cite{Marcolli:2009in} where Majorana mass terms for right-handed
neutrinos were included and the see-saw mechanism was taken into
account.  In our renormalization group analysis of the Higgs
potential, following Ref.~\cite{Espinosa:2007qp}, the choice of
boundary conditions is the standard one motivated by particle physics
considerations. The focus here has of course been on the different
boundary conditions at low energies for which a flat section develops
in the Higgs potential.

The relations above rely on the validity of the asymptotic expansion
at $\Lambda$, and are therefore tied intimately to the scale at which
the expansion is performed.  There is no a priori reason for the
constraints to hold at scales below $\Lambda$ --- they represent mere
boundary conditions. The constraint $\xi(\Lambda)=1/12$ by itself
therefore does not require the coupling to remain conformal all the
way down to present energy scales, or even during an inflationary
epoch, since it may run with the energy scale. However, we will assume
no running in $\xi$ as the arguments laid out ({\sl
  see}, discussion in Section \ref{slow-roll}A) above still apply.

As we can see from the results presented above, the conformally
coupled Higgs field in the Spectral Action Standard Model is not a
viable candidate for inflaton if the coupling remains conformal at all
scales. However, at present it is still unclear whether
conformal~\footnote{The coupling term between the Higgs field and the
  Ricci curvature, appearing in the spectral action functional, is
  $-f_0/(12\pi^2)aR|\phi|^2$, which after rescaling ${\bf
    H}=(\sqrt{af_0}/\pi)\phi$, leads to the term $-R|{\bf
    H}|^2/12$. This indeed shows the conformal coupling between the
  background and the Higgs field.}  invariance and $\xi=1/12$ is a
generic feature of models from noncommutative geometry. If it turns
out not to be, one can proceed along the line of the analyses
presented in Refs.~\cite{Nelson:2009wr,Marcolli:2009in}.

\subsection{Inflation through the massless scalar field}
The spectral action gives rise to an additional massless scalar
field~\footnote{The field $\sigma$ is unlike all other fields in the
  theory, such as the Higgs field and gauge fields. Usually one starts
  with a parameter in the Dirac operator of the discrete space, and
  then inner fluctuations of the product space would generate the
  dynamical fields. The only exception being the matrix entry that
  gives mass to the right-handed neutrinos, where the parameter can
  either remain as such, or one can use the freedom to make it a
  dynamical field, which {\sl a priori} may lead to important
  cosmological consequences~\cite{Ali}. Note that the $\sigma$ field
  was not considered in the original noncommutative geometry spectral
  action analysis presented in Ref.~\cite{Chamseddine:2006ep}, where
  the authors were mainly interested in recovering the Standard
  Model.}~\cite{Chamseddine:2009yf}, denoted by $\sigma$. Including
this field, the cosmologically relevant terms in the Wick rotated
action read
\be
\begin{split}
S=\int\,d^4x\sqrt{-g}\left\{\frac{1}{2\kappa^2}
R - \xi_{\bf H} R{\bf H}^2 - \xi_\sigma R \sigma^2 \right.\\ 
\left.-\frac{1}{2}(\nabla {\bf H})^2  -\frac{1}{2}(\nabla \sigma)^2 
- V({\bf H},\sigma)\right\}
\end{split}
\ee
where
\be
V({\bf H},\sigma)=\lambda_{\bf H}{\bf H}^4-\mu_{\bf H}^2{\bf H}^2+\lambda_\sigma
\sigma^4+\lambda_{{\bf H}\sigma}|{\bf H}|^2\sigma^2~.
\ee
The constants are related to the underlying parameters as
follows~\footnote{Note that a similar action has been studied in
  Ref~\cite{Lerner:2009xg}, but in their analysis the additional
  scalar field has a nonzero mass and the nonminimal couplings are
  studied in the previously mentioned large negative $\xi$ regime,
  which flattens the classical quartic potential in the Einstein
  frame.}:
\begin{align}
\xi_{\bf H} &=\frac{1}{12}~~~~~~~~~~~~~~,  &\xi_\sigma &=\frac{1}{12}\\
\lambda_{\bf H} &=\frac{\pi^2\mathfrak{b}}{2f_0\mathfrak{a}^2}~~~~~~~~~~, 
&\lambda_\sigma &= \frac{\pi^2\mathfrak{d}}{f_0\mathfrak{c}^2}\\
\mu_{\bf H} &=2\Lambda^2\frac{f_2}{f_0}~~~~~~~~~~,
&\lambda_{{\bf H}\sigma}&=\frac{2\pi^2\mathfrak{e}}{a\mathfrak{c}f_0}~.
\end{align}

This action also admits a rescaling of the metric which transforms it
to the Einstein frame. The rescaled metric $\hat
g_{\mu\nu}=f({\bf H},\sigma)g_{\mu\nu}$ with
$f({\bf H},\sigma)=1-2\xi_{\bf H}{\bf H}^2-2\xi_\sigma\sigma^2$ is now
accompanied by the new fields $\chi_{\bf H}$ and $\chi_\sigma$ related to
the Jordan frame fields by
\begin{align}
\frac{d\chi_{\bf H}}{d{\bf H}}
&=\frac{\sqrt{1-2\xi_{\bf H}(1-12\xi_{\bf H}){\bf H}^2}}{f({\bf H},\sigma)}
\nonumber\\
 &\xrightarrow{\text{CC}}\ \ \frac1{f({\bf H},\sigma)}~,
\\ \frac{d\chi_\sigma}{d\sigma}
&=\frac{\sqrt{1-2\xi_\sigma(1-12\xi_\sigma)\sigma^2}}{f({\bf H},\sigma)}
\nonumber\\ &\xrightarrow{\text{CC}}\ \ \frac1{f({\bf H},\sigma)}~.
\end{align}
The Einstein frame Lagrangian reads
\begin{align}
S&=\int\,d^4x\sqrt{-\hat g}\left\{\frac{1}{2\kappa^2}\hat R
-\frac{1}{2}(\hat\nabla \chi_{\bf H})^2- \frac{1}{2}(\hat\nabla
\chi_\sigma)^2\right. \nonumber\\ &\quad
\ -\left.P(\chi_{\bf H},\chi_\sigma)
\hat\nabla_\mu\chi_{\bf H}\hat\nabla^\mu\chi_\sigma
- V(\chi_{\bf H},\chi_\sigma) \vphantom{\frac{1}{2}(\hat\nabla
  \chi_{\bf H})^2}\right\},
\end{align}
with 
\be \hat V(\chi_{\bf H},\chi_\sigma)=\frac{V({\bf H},\sigma)}{f({\bf
    H},\sigma)^2}~, \ee
and a novel coupling 
\begin{align}
P(\chi_{\bf H},\chi_\sigma)&=\frac{24\kappa^2\xi_{\bf H}\xi_\sigma}
{f({\bf H},\sigma)^2} \frac{d{\bf H}}{d\chi_{\bf
    H}}\frac{d\sigma}{d\chi_\sigma}{\bf H}\sigma\nonumber\\ &
\ \ \xrightarrow{\text{CC}}\ \ \frac{\kappa^2}{6}\sigma{\bf H}~.
\end{align}
Note that there exists no conformal transformation which gets rid of
both the nonminimal coupling to gravity and the cross-term
$P(\chi_{\bf H},\chi_\sigma)$~\cite{Kaiser:2010ps}. However, at conformal
coupling $P(\chi_{\bf H},\chi_\sigma)$ can be neglected as long as
$\sigma{\bf H}\ll6\kappa^2$. We are then left with a minimally coupled
theory of two scalar fields with potential $\hat
V(\chi_{\bf H},\chi_\sigma)$. When the mass term is negligible the theory
is symmetric in the two fields.

Consider the first slow-roll parameter for the $\sigma$-field, defined as
\begin{align}
\hat \epsilon_\sigma &=\frac{1}{2\kappa^2}\frac{1}{\hat V^2}
\left(\frac{\partial\hat V}{\partial\sigma}\right)^2
\left(\frac{\partial\chi_\sigma}{\partial\sigma}\right)^{-2}\\ 
&=\frac{1}{2\kappa^2}
     [\lambda_{\bf H}{\bf H}^4+\lambda_\sigma \sigma^4+\lambda_{{\bf
           H} \sigma}|{\bf H}|^2 \sigma^2]
     ^{-2}\nonumber\\ &~~~~\times\sigma^2
     \left[\left(4\lambda_\sigma\sigma^2+2\lambda_{{\bf H}\sigma}{\bf
         H}^2\right) f({\bf H},\sigma)
       \vphantom{\frac23\kappa^2}\right.\nonumber\\ 
&~~~~~~~~~~~~\left.+\frac23\kappa^2\left(\lambda_\sigma\sigma^4+
       \lambda_{\bf H}{\bf H}^4+\lambda_{{\bf H}\sigma}{\bf
         H}^2\sigma^2\right)\right]^2~.
\end{align}
For ${\bf H}=0$ this reduces to the earlier case and one gets an
insufficient numebr of $e$-folds below the Planck scale. If we have a
nonzero ${\bf H}$ however, say ${\bf H}\sim\kappa^{-1}$ close to the Planck
mass, then the situation changes somewhat. Due to the additional terms
in $\hat\epsilon_\sigma$, the coupling constants do not fall out of
the expression, and they can therefore influence the magnitude of the
integrand in the number of $e$-folds. For this effect to take place
however, it is necessary that the assisting field maintains a
relatively large value throughout the inflationary era driven by the
inflaton. This in turn requires the curvature of the potential to be
much less in the direction of the constant field. Since only quartic
terms arise in the model, the quartic self-coupling of the assisting
field is then required to be much lower than that of the inflaton. But
in that case, the new terms due to the assisting field are not large
enough to enable the inflaton to generate sufficient number of
$e$-folds.  The situation is of course entirely symmetric in the two
fields (except for the nonzero ${\bf H}$-mass which is negligible at high
energies), so the r\^oles of the two fields may well be interchanged
depending on which constraints lay on the respective coupling
constants.

\section{Running of the Gravitational constant}
\label{running-gravitational}
At the scale $\Lambda$, the gravitational constant is related to the
geometric parameters of the theory by
\be
\kappa^2=\frac{12\pi^2}{96f_2\Lambda^2-f_0\mathfrak{c}}~;
\ee 
$f_0$ is fixed by
one of the unification conditions
\be f_0=\frac{\pi^2}{2g_1^2}\ \ , \ \ f_0=\frac{\pi^2}{2g_2^2}\ \ ,
\ \ f_0={\pi^2}{2g_3^2}\nonumber~, \ee
$f_2$ is an unconstrained parameter in $\mathbb{R}^*_+$, and
$\mathfrak{c}$ is determined by the renormalization group equations.
Note that this value of the gravitational constant does not need to be
the same as its present value,
$\kappa^2=[2.43\times10^{18}\text{GeV}]^{-2}$, since the gravitational
constant may run.

Indeed, such a running has been suggested~\cite{Marcolli:2009in} due
to the relation between $\kappa^2$ and $\mathfrak{c}={\rm
  Tr}(MM^\dagger)$; $M$ stands for the Majorana mass term. The
coefficient $\mathfrak{c}$ is a function of the neutrino mass matrix
subject to running with the renormalization group equations dictated
by the particle physics content of the model, in this case the
Standard Model with additional right-handed neutrinos with Majorana
mass terms.  Since the renormalization group flow runs between a
unification energy $\Lambda$, taken to be of the order of $2\times
10^{16}$ GeV, down to the electroweak scale of 100 GeV, the parameter
$\mathfrak{c}$ runs as a function of $\Lambda$, with assigned initial
conditions at the preferential energy scale of unification. One may
thus deduce that through the running of $\kappa^2$, the number $N$ of
$e$-folds may increase. However, since at conformal coupling $N$ is a
logarithmic function of $\kappa^2$, the gravitational constant would
have to change drastically, in order for $N$ to have an interesting
change. As previously mentioned, we need a very significant running of
the gravitational constant in order to get inflation --- a
\emph{local} kind of inflation where the flat region is confined to a
small range of energies where the potential is flattened.

In conclusion, unless modification of the spectral triple allows for a
nonconformal boundary value of $\xi$, there seems to be no viable
slow-roll scenario for any of the two scalars. Furthermore, if one
assumes the validity of the suggestion in Ref.~\cite{Marcolli:2009in}
in relating the running of $\mathfrak{c}$ and $\kappa^2$, the only
situation in which this could trigger inflation (with conformal
coupling) would be one in which the running changes drastically, {\sl
  e.g.}, through the see-saw mechanism. However, the inevitable lack
of differentiability of the renormalized couplings at see-saw
scales~\cite{Marcolli:2009in,Antusch:2005gp} makes such a scenario
very unlikely and also inaccessible to slow-roll analysis.  \\

\section{Conclusions}
\label{conclusions}
In many realistic cosmological models, the nonminimal coupling of the
scalar field to the Ricci curvature cannot be avoided. In particular,
there are arguments requiring a conformal coupling between the scalar
field and the background curvature. The existence of such a term will
generically lead to difficulties in achieving a slow-roll inflationary
era. In this paper, we have investigated whether two-loop corrections
to the Higgs potential could lead to a slow-roll inflationary period
in agreement with the constraints imposed by the CMB measurements. Our
findings do not favor the realization of such an era. More precisely,
even though slow-roll inflation can be realized, we cannot satisfy the
COBE normalization constraint for any values of the top quark and
Higgs masses allowed from current experimental data.

We have in particular investigated Higgs inflation in the context of
the Noncommutative Geometry Spectral Action, which provides an elegant
explanation for the phenomenology of the Standard Model. Within this
context, a conformal coupling arises naturally between the Higgs field
and the Ricci curvature.  It is also important to note that once
conformal coupling is set at the preferential (boundary) energy scale
of the spectral action model, then it will remain conformal at all
scales.  Running of the gravitational constant and corrections by
considering the more appropriate de\,Sitter, instead of a Minkowski,
background do not favor the realization of a successful inflationary
era.  The NCG Spectral Action provides in addition to the Higgs field,
another (massless) scalar field which exhibits no coupling to the
matter sector. Our analysis has shown that neither this field can lead
to a {\sl successful} slow-roll inflationary era if the coupling
values are conformal. One may be able to improve upon this (negative)
conclusion, if important deviations of $\xi$ from its conformal value
can be allowed; the value $\xi=1/12$ may turn out that is not a
generic feature of NCG models.

\bigskip
\section{Acknowledgments}
This work is partially supported by the European Union through the
Marie Curie Research and Training Network {\sl UniverseNet}
(MRTN-CT-2006-035863).

\bibliography{inflation}{} \bibliographystyle{unsrt}

\end{document}